\documentclass[aps,prc,twocolumn,showpacs,superscriptaddress,
preprintnumbers]{revtex4-1}

\usepackage{graphicx}
\usepackage{amsmath}
\usepackage[dvipsnames]{xcolor}
\usepackage{enumitem}
\usepackage{hyperref}

\hypersetup{ 
    pdfnewwindow=true,      
    colorlinks=true,       
    linkcolor=blue,          
    citecolor=blue,        
    filecolor=blue,      
    urlcolor=blue        
}  

\newcommand{\be}{\begin{eqnarray}}
\newcommand{\ee}{\end{eqnarray}}

\begin{document}

\title{Model Selection for Pion Photoproduction}

\author{J.~Landay}
\email{jlanday@gwmail.gwu.edu}
\affiliation{The George Washington University, Washington, DC 20052, USA}

\author{M.~D\"oring}
\email{doring@gwu.edu}
\affiliation{
Institute for Nuclear Studies (INS); Astronomy, Physics and Statistics Institute of Sciences (APSIS), 
The George Washington University, Washington, DC 20052, USA}
\affiliation{Theory Center, Thomas Jefferson National Accelerator Facility, Newport News, VA 23606, USA}

\author{C.~Fern\'andez-Ram\'{\i}rez}
\email{cesar.fernandez@nucleares.unam.mx}
\affiliation{Instituto de Ciencias Nucleares, Universidad Nacional Aut\'onoma de M\'exico, Ciudad de M\'exico 04510, Mexico}

\author{B.~Hu}
\affiliation{The George Washington University, Washington, DC 20052, USA}

\author{R.~Molina}
\affiliation{The George Washington University, Washington, DC 20052, USA}


\preprint{JLAB-THY-16-2364}

\begin{abstract}
Partial-wave analysis of meson and photon-induced reactions is needed to enable the comparison of many theoretical
approaches to data. In both energy-dependent and independent parametrizations of partial waves, the selection
of the model amplitude is crucial. Principles of the $S$-matrix are implemented to different degree in different approaches;
but a many times overlooked aspect concerns the selection of undetermined coefficients and functional forms for fitting, 
leading to a minimal yet sufficient parametrization.
We present an analysis of low-energy neutral pion photoproduction using the 
Least Absolute Shrinkage and Selection Operator (LASSO)
in combination with criteria from information theory 
and $K$-fold cross validation. These methods are not yet widely known
in the analysis of excited hadrons but 
will become relevant in the era of precision spectroscopy.
The principle is first illustrated with synthetic data; then, its feasibility for real data is demonstrated 
by analyzing the latest available measurements of differential cross sections ($d\sigma/d\Omega$), 
photon-beam asymmetries ($\Sigma$), and target asymmetry differential cross sections
($d\sigma_T/d\equiv T\,d\sigma/d\Omega$) in the low-energy regime.
\end{abstract}


\maketitle

\section{Introduction}
The understanding of the strong interaction in the hadronic energy regime is an important unresolved issue that 
has regained a lot of attention in the last years due to the advances in detection techniques, accelerator technologies,
first principle Quantum Chromodynamics (QCD) analyses, 
and S-matrix theory amplitude analysis techniques \cite{ATHOS,Briceno2016, Dudek2016, Lang:2016hnn, cfr2016,Fernandez-Ramirez:2015tfa}.
These developments have lead to a broad effort to 
build and perform experiments that are or will be collecting an unprecedented amount of
data on hadron reactions, e.g.,
BELLEII \cite{BELLE2a,BELLE2b},
BESIII  \cite{BEPC},
CLAS12 \cite{CLAS}, 
CMS \cite{CMS},
COMPASS \cite{COMPASS},
ELSA \cite{ELSA,ELSA2},
GlueX \cite{GlueX},
J-PARC \cite{JPARC}, 
KLOE2 \cite{KLOE2}, 
LHCb \cite{LHCb}, 
MAMI \cite{MAMI}
and PANDA \cite{PANDA}.

Partial-wave analysis of hadronic reactions is a prerequisite for many theoretical approaches to access information from the
experimental data, especially if the comparison to hadron resonances is the goal.
Narrowing the focus to photoproduction reactions, their decomposition into 
partial waves (multipoles) is usually performed through an
energy-dependent (ED) parametrization of the amplitude. 
As long as data are not abundant and precise enough, it is not yet possible to perform
a (truncated partial-wave) complete experiment~\cite{Wunderlich:2014xya, Workman:2016irf} in the resonance region. 
A parametrization in energy is needed for the determination of resonances, or as a stabilizing 
starting point for single-energy (SE) solutions in which energy-binned data are fitted independently.
For both ED and SE analyses, the selection of fit parameters is a fundamental 
problem that we address in this study for the case of neutral pion photoproduction in the low energy region.
We use this well-studied reaction as a benchmark for various techniques and as a template for future works because of 
the well-established theoretical framework and the availability of high-quality data 
which allow an (almost) model-independent SE extraction~\cite{Hornidge:2012ca,Fernandez-Ramirez:2014ffa}.

In the analysis of photoproduction experiments, 
the parametrization of the amplitude is chosen according to the considered energy range. 
For low-energy neutral pion photoproduction,
Heavy Baryon Chiral Perturbation Theory 
(HBChPT)~\cite{Bernard:1996ft,Bernard:1995cj,Bernard:1994gm,Bernard:1994ds,Bernard:1993bq,Bernard:1992nc,Bernard:1991rt,Bernard:2005dj,FernandezRamirez:2012nw}
and Relativistic Baryon Chiral Perturbation Theory (RBChPT)~\cite{Hilt:2013uf}
provide effective parametrizations.
Both deliver equally good descriptions of the latest experimental data up to $E_\gamma \simeq 170$ MeV 
in the laboratory frame~\cite{Hornidge:2012ca,Schumann:2015ypa}.
Polynomial parametrizations which incorporate unitarity in the S wave have also proved to be an excellent description
of the data up to $E_\gamma \simeq 185$ MeV where the $\Delta(1232)$ contribution begins to be relevant
\cite{FernandezRamirez:2009jb,Hornidge:2012ca,Fernandez-Ramirez:2013qqa}.
ChPT calculations including isospin breaking have been performed in 
Refs.~\cite{Baru:2010xn,Hoferichter:2009ez, Baru:2011bw} and
the inclusion of the $\Delta(1232)$ resonance as an explicit degree of freedom in RBChPT
allows to extend the agreement between theory and data up to $E_\gamma \simeq 200$ MeV~\cite{Blin:2014rpa, Blin:2016itn}.
The RBChPT calculation in~\cite{Hilt:2013uf} has been extended to include also electroproduction of
charged pions~\cite{Hilt:2013fda}. 

In an explicit parametrization of partial waves one cannot incorporate an infinite number (the series has to be truncated)
and we need to determine how many terms one needs to incorporate in order 
to provide an accurate, yet minimally parameterized, description of the physics involved.
In Refs.~\cite{FernandezRamirez:2009su,FernandezRamirez:2009jb} 
it was determined that in the low-energy neutral pion photoproduction
region we need to incorporate up to $D$ waves and that higher partial waves can be safely dismissed.
Although $D$ waves are not necessary to describe the experimental data, not including them prevents the accurate
extraction of the $S$ wave, which is small since it vanishes in the chiral limit and provides insight into chiral symmetry breaking
\cite{Bernstein:1998ip}.
The structure of the observables in terms of the multipoles can be found in \cite{multipoles1,multipoles2,FernandezRamirez:2009jb}.

Beyond energies in which a systematic treatment is possible, effective field theory 
approaches~\cite{Ronchen:2014cna, Ronchen:2015vfa, Kamano:2013iva, Huang:2011as, Tiator:2010rp,FernandezRamirez:2005iv,FernandezRamirez:2008,Gasparyan:2010xz,JuliaDiaz:2007fa}, 
$K$-matrix parametrizations, or related approaches are used in~\cite{Workman:2012jf, Anisovich:2011fc, Drechsel:2007if}. 
At high energies, Regge parametrizations are very effective~\cite{Sibirtsev:2009bj, Mathieu:2015eia}. 
Formulations to provide amplitudes that cover the entire energy region from threshold to the highest energies 
are under development~\cite{Mathieu:2015gxa, Nys:2016vjz}.
Yet, sometimes partial waves are parameterized purely phenomenologically in terms of functions that
are in agreement with
basic $S$-matrix principles such as coupled-channel two-body unitarity, the correct threshold behavior 
or Fermi-Watson's theorem \cite{Watson,Fermi}, 
but that are otherwise left free to ensure a high degree of model independence as in the SAID approach~\cite{Workman:2012jf}. 

In general, new high-precision polarization data indeed 
lead to more consistent multipole solutions among different analysis groups
although discrepancies remain~\cite{Beck:2016hcy}. A step towards the goal of matching solutions has been done recently. 
Providing the necessary information for other groups
to carry out correlated $\chi^2$ fits of $\pi N$ partial waves, 
the statistical influence of elastic pion-nucleon scattering has been quantified~\cite{Doring:2016snk}.
This gives a more statistical foundation for multi-reaction analyses 
by many groups in the search for new excited baryons. 
Complementary information from hadron beams could also lead to more consistent solutions among the 
different partial-wave analysis approaches~\cite{Briscoe:2015qia}.
Considerations of which observables and which precision are necessary to discriminate 
models are also a necessary step forward to find definite answers in baryon spectroscopy~\cite{Nys:2016uel}.

The selection of fit parameters in most of these approaches is important. 
If the amplitude is under-parameterized, the quality of the data description 
is not satisfactory and the quality of the extracted amplitudes is difficult to assess.
Over-parametrization can result in limited predictability of the amplitude outside the fitted data range and inflated uncertainties. 
Furthermore, problems in the data themselves 
(incompatibility of data, systematics, or even statistics) may be interpreted as significant physics in over-parameterized fits.

Most notably, in many approaches resonances are introduced 
in the parametrization as explicit terms, that will unavoidably improve the
fit quality at the cost of potentially false positive resonance signals~\cite{Ceci:2011ae}. 
To control this problem, groups use mass scan techniques in which, 
ideally, a minimum of the $\chi^2$ as a function of the resonance mass appears 
in more than one analyzed channel~\cite{Arndt:2003ga}. In the SAID approach,
resonances appear dynamically generated, meaning that poles in the amplitude 
can appear without manual intervention, if required by data~\cite{Workman:2012jf}. 
In Refs.~\cite{DeCruz:2011xi, DeCruz:2012bv}, 
the most probable resonance content (Bayesian evidence) is determined considering kaon photoproduction. 
Bayesian priors have also been used to restrict the low-energy constants  in effective field theories to 
natural values and estimate the truncation errors~\cite{Wesolowski:2015fqa, Furnstahl:2015rha}.

If a flexible background with resonance terms on top of it is provided, 
the task consists in minimizing the number of resonances and 
only accept them as physically significant if the background cannot provide a satisfactory description. 
The  Least Absolute Shrinkage and Selection Operator (LASSO)~\cite{orilasso, Lasso, ISL} provides a tool to scan 
a plethora of different models, in particular multiple combinations of different resonances. 
Manually, such a scan would be impossible
due to the large number of combinations, but the LASSO provides an automatized, 
``blind-folded'' technique~\cite{Guegan:2015mea}. 

Having the above-mentioned extensions for future work in mind, we concentrate in this study on the question of how to select 
the simplest amplitude for a real-life example of photoproduction reactions. We choose low-energy neutral-pion photoproduction, 
$\gamma p\to\pi^0p$, for which data of unprecedented precision exist from 
the same experimental setup at MAMI \cite{Hornidge:2012ca,Schumann:2015ypa}, 
thus minimizing the potential impact from conflicting data or systematic uncertainties. 
The differential cross section $d\sigma/d\Omega$, photon beam asymmetry $\Sigma$,
and target polarization differential cross section $d\sigma_T/d\Omega\equiv T\,d\sigma/d\Omega$ are analyzed. 
The entire presentation is kept as pedagogical as possible, 
even quoting textbook formulae for easier reference. 
The most relevant references on statistical analysis are \cite{Lasso, ISL, freund}.

\section{Formalism} \label{sec:formalism}

\subsection{Parametrization}
\label{sec:parametrization}
An energy-dependent parametrization  is formulated for both real 
and imaginary parts of the three $P$ waves, as well as for the real parts of $E_{0+}$ and the four $D$-wave multipoles,
\be
{\rm Re, Im}\, {\cal M}_{L\pm}&=&\frac{q_{\pi^0}^\ell}{m_{\pi^+}^{\ell+1}}\,
\sum_{i=0}^{i_{\rm max}} \frac{a_i}{10^{-i}} \left(\frac{\omega_{\pi_0}-m_{\pi_0}}{m_{\pi^+}}\right)^i \ 
\label{par1}
\ee
where $q_{\pi^0}$ 
is the center-of-mass momentum of the neutral pion, $\omega_{\pi_0}^2=m_{\pi^0}^2+q_{\pi^0}^2$, and
$a_i$ are the fit parameters.
The quantity ${\cal M}$ stands for the electric ($E_{L\pm}$) and magnetic ($M_{L\pm}$) multipoles, 
or alternatively, the partial waves $P_1$, $P_2$ and $P_3$ for the $P$ waves,
related to each other by
\begin{equation}
\begin{split}
E_{1+}=& \frac{1}{6} (P_1 +P_2) \\
M_{1+}=&\frac{1}{6} (P_1-P_2+2\, P_3)  \\
M_{1-}=&\frac{1}{3} (P_3+P_2-P_1)\ .
\end{split}
\end{equation}
The parametrization in energy consists of a factor providing the correct threshold behavior 
and a Taylor expansion in momentum-squared around the neutral pion threshold.
This expansion is theoretically justified by ChPT calculations~\cite{Bernard:1996ft,Bernard:1995cj,Bernard:1994gm,Bernard:1994ds,Bernard:1993bq,Bernard:1992nc,Bernard:1991rt,Bernard:2005dj}.
In principle, it is preferable to use a set of polynomials in energy that is orthogonal in the fitted energy window, 
which reduces the correlations among parameters. 
However, in this particular example we could not observe any improvement when doing so.

Furthermore, in Eq.~(\ref{par1}) for the real parts of the multipoles, one has $\ell=L$, 
while for the imaginary parts of the $P$-wave multipoles, $\ell=3L+1$=4 as can be obtained from Watson's theorem.
We do not provide any imaginary parts for the $D$-waves because, 
on one hand, they are extremely small and, on the other end, restricting them to be real fixes the overall-phase ambiguity.
Fit parameters are generically called $a_i$ throughout, and we omit the indices specifying 
to which partial wave (real or imaginary part) they belong. 
The  same applies to the cut-off $i_{\rm max}$. 
For the real parts of the $P$-waves and $D$-waves, $i_{\rm max}=4$ 
while for the imaginary parts of the $P$-wave, $i_{\rm max}=0$. 
Other factors in Eq.~(\ref{par1}) serve to make all parameters dimensionless and of natural size.
For the $S$-wave multipole $E_{0+}$, a real-valued term of the form of Eq.~(\ref{par1}) with $i_{\rm max}=4$ 
is supplemented by a term of the form
\begin{equation}
\Delta E_{0+}=i \frac{q_{\pi^+}}{m_{\pi^+}^2}  \sum_{i=0}^{i=2} \frac{a_i}{10^{-i}}  \left(\frac{q_{\pi^+}}{m_{\pi^+}}\right)^{2i} 
\label{eq:ime0p}
\end{equation}
to take into account the $\pi^+n$ threshold cusp. 
The term provides an imaginary part above the $\pi^+n$ threshold and contributes to the real part of $E_{0+}$ below it. 
In total, there are $i_{\rm max}=46$ free fit parameters. 
The number of available parameters is simply chosen such that every multipole can be grossly over-fitted. 

\subsection{Criteria from information theory}
With this parametrization at hand, we turn to the LASSO method to select the 
\textit{simplest} model~\cite{orilasso,Lasso, ISL}. The penalized $\chi_T^2$ is defined as follows:
\be
\chi^2_T(\lambda)=\chi^2(\lambda)+P(\lambda),
\label{totalchi}
\ee
with 
\be
P(\lambda)=\lambda^4\sum_{i=1}^{i_{\rm max}} |a_i| \ .
\label{penalty}
\ee
In practice, one scans an entire range of $\lambda$, continuously minimizing $\chi_T^2$. 
From there, we turn to various criterions from information theory in order to determine the optimal $\lambda$. 
Note that the power of four in Eq.~(\ref{penalty}) is simply chosen to provide a more convenient graphical representation 
of these criteria in the following plots. 
The three criteria that we use to find an optimal $\lambda$ are the 
Akaike Information Criterion (AIC)~\cite{AIC, AIC2}, 
a finite sample size corrected version of the AIC (AICc)~\cite{AICc}, 
and the Bayesian Information Criterion (BIC)~\cite{BIC}. 
The three are defined as
\begin{equation}
\begin{split}
\text{AIC} =& 2k -2\log(L) = 2k + \chi^2 \ ,  \\
\text{AICc} =& \text{AIC} + \frac{2k(k+1)}{n-k-1} \ ,  \\
\text{BIC} =& k\log(n) -2\log(L)= k\log(n) + \chi^2 \ ,
\end{split}
\end{equation}
where $k$ is the number of parameters which changes dynamically as a function of $\lambda$, 
$n$ is the number of data points, and $L$ is the likelihood. 
We define the number of degrees of freedom (d.o.f.) for each fit as ${\rm d.o.f.}=n-k$.
For all three of the criteria, the optimal value of $\lambda$ is given by the respective minimum. 
This is because the AIC and the BIC take on small values for models with low test error. 
Assuming a Gaussian model, the BIC is proportional to the AIC, however the BIC tends to penalize models 
with more parameters due to the factor $\log (n)$ which allows for a more distinct minimum to be seen 
and a better indication of which model to use. 
For a further comparison between the AIC, AICc and the BIC, see Refs.~\cite{Lasso, ISL}. 

\begin{figure}
\begin{center}
\includegraphics[width=0.9\linewidth]{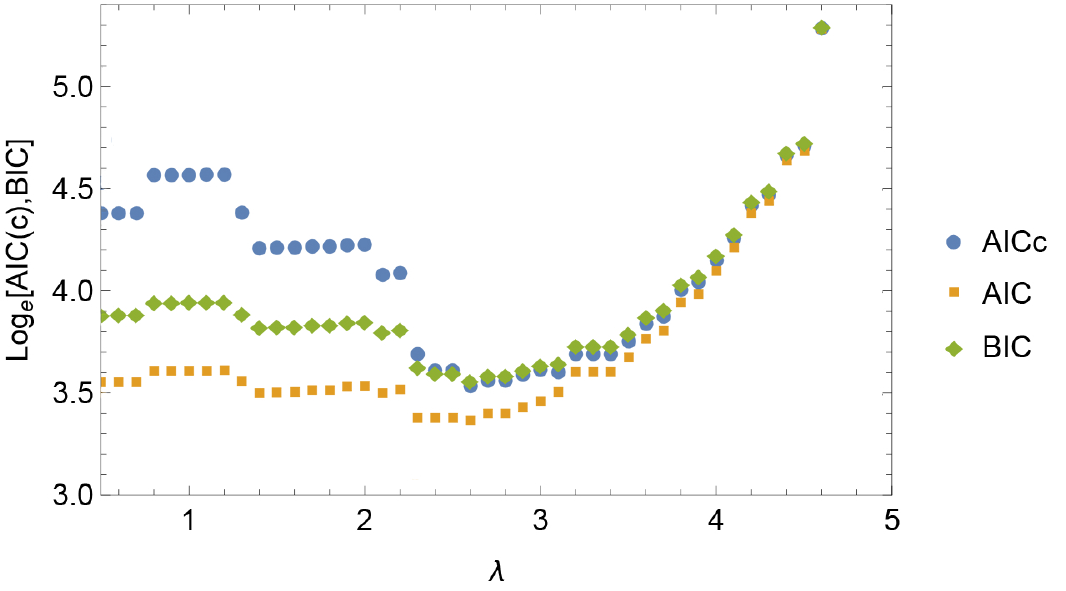}
\end{center}
\caption{The AIC, AICc, and BIC criteria for a simple $\chi^2$ fit to 22 data (see text). 
All three methods exhibit the minimum at the same value of $\lambda$ indicating the optimal, 
i.e., simplest model. The correction term for a finite number of data in the AICc indeed produces 
a more pronounced minimum than for the AIC. The discontinuities occur every time a fit parameter 
is set to zero by LASSO.}
\label{fig:criteria1}
\end{figure}

In Fig.~\ref{fig:criteria1} we show the different criteria as a function of the penalty $\lambda$ 
in a simple simulation of fitting 22 synthetic data generated from a low-order polynomial 
with a model that includes that low-order polynomial as solution and also allows for over-fitting. 
One recognizes here a clear difference between the AIC and the AICc. 
The correction for the finite number of data points indeed leads to a larger AICc at small $\lambda$, 
providing a better identification of the minimum than the AIC. 
In the case of pion photoproduction considered later, the number of data is so large, 
though, that the difference between AIC and AICc becomes irrelevant.

%

\subsection{Cross validation}
Another method to find a model with minimal parametrization yet with accurate description of data is cross
validation~\cite{orilasso, Lasso, ISL}. In short, data are randomly divided into a training set and a validation set. 
For a given $\lambda$, the penalized $\chi_T^2$ of the training set is minimized and the $\chi^2_V$ of the validation set is determined
from that fit (without refitting). While $\chi^2_T$ is clearly a monotonously increasing function of $\lambda$, 
the situation is different for $\chi^2_V$. For very large $\lambda$, $\chi^2_V$ is large as well, because 
the data are under-fitted. However, as $\lambda$ decreases a point is reached below which $\chi^2_T$ is over-fitted, i.e.,
non-physical structures such as statistical fluctuations in the training set are described. These fluctuations are different in 
the validation set, such that the validation $\chi^2_V$ becomes {\it larger} again as $\lambda$ decreases further.
The minimum in $\chi^2_V$ is then regarded as the 
\textit{sweet spot} for $\lambda$, i.e., the point where the fit optimally describes the data
without describing fluctuations. 

The method has been recently applied in \cite{Sato:2016tuz} for the determination 
of parton distribution functions. Another example is given in \cite{Agadjanov:2016mao}. There, the task 
consisted in effectively smoothing a function that is subject to oscillations from unphysical finite-volume effects. 
In that example, the unphysical effect was not given by fluctuations which demonstrates that LASSO in combination with 
cross validation is a method with broader range of applicability than needed here. 

The minimum of $\chi^2_V$ itself carries uncertainties. In practical terms it is numerically demanding to carry out the above-mentioned
separation of training and validation sets for all possible combinations of data. An approximate method is given by $K$-fold
cross validation in which the (uncorrelated) data are randomly divided in a few (here: 5) partitions. In five different cross validation
runs, four of the five partitions serve as training set while the fifth serves as validation set. The different outcomes are used
to estimate the uncertainty of $\chi^2_V$ at each $\lambda$. 
One can then further constrain the search for the simplest model by selecting the 
$\lambda_{\rm opt}>\lambda_{\rm min}$ that is compatible within errors 
with the minimum of $\chi^2_V$ at $\lambda=\lambda_{\rm min}$. In practice, 
this means to search for the $\lambda_{\rm opt}$ 
at which $\chi^2_V(\lambda_{\rm opt})=\chi_V^2(\lambda_{\rm min})+\Delta$ where $\Delta$ is the uncertainty 
of $\chi^2_V$ at $\lambda=\lambda_{\rm min}$. This is referred to as the 1-$\sigma$ rule~\cite{one-sigma1, one-sigma2}.

\subsection{Bootstrap and non-Gaussian uncertainties}
\label{sec:bootstrap}
Although it is common knowledge, we briefly mention the bootstrap technique to keep the presentation as pedagogical as possible, 
and we specify the way bootstrapping is implemented in this study. 
For a very similar procedure, see Refs.~\cite{Fernandez-Ramirez:2015tfa,Blin:2016dlf}. 
Once a model is selected, the propagation of uncertainties from data to results
can be carried out using bootstrap. Here, the results are 
given by the multipoles and the cusp parameter 
at threshold, $\beta_0$ (see Sec.~\ref{sec:realdatadiscussion} for discussion). 
This resampling technique
allows to trace non-linear uncertainties and is, thus, in principle, superior to methods using the covariance matrix.
Here, we repeatedly perform complete random resamplings of all data points according 
to their uncertainties and then refit each resampled data set.
From each set of the resulting fit parameters, the multipoles and the cusp parameter $\beta_0$ are evaluated. 

If the resulting distribution, e.g., of a multipole at fixed energy $W$, 
is Gaussian one can just estimate the variance and determine the final
uncertainty by its square root. If the distribution is very skewed it is more meaningful 
to determine the 68\% confidence level (CL) interval
by cutting off the 16\% largest and 16\% smallest values. 

This leads to a related comment concerning the fitting of the beam asymmetry $\Sigma$ that is one of the 
considered observables in this study. Usually, the statistical uncertainty 
in polarization observables provided by experiment is treated as Gaussian
in partial wave analysis, as if it originated from the measurement 
of a cross section in the limit of many counts. However, these observables $O$ are ratios
of the difference of positive Poisson distributions divided by their sum, 
i.e., they are restricted to $|O|< 1$. Thus, strictly speaking, those uncertainties cannot be regarded as Gaussian
and maximizing the likelihood cannot be achieved by minimizing the $\chi^2$.
The bias is maximal for $|O|\approx 1$. The size of the beam asymmetry $\Sigma$ is far from this limit at the 
low energies considered in this study and we neglect the bias here.

\section{Results}

\subsection{LASSO in a Benchmark Model}

For a controlled test of the discussed methods, and before dealing with experimental data, 
we test our ideas with synthetic data. In doing so we proceed in the following way:

{\bf 1.} We generate synthetic data from a given set of multipoles and study to which precision and accuracy 
we can reconstruct that known set. 
To that end we first build a benchmark model ($\mathcal{B}$-model) which is a reduced version of the one described 
by Eqs. (\ref{par1}) and (\ref{eq:ime0p}). 
We build it setting all the $a_i$ to zero except for: $a_0$ and $a_1$
for the real parts of every $S$ and $P$ wave in Eq.~(\ref{par1}) and $a_0$ in Eq.~(\ref{eq:ime0p}),
totaling 9 parameters, i.e. 2 for each $P$-wave multipole and 3 for the $S$ wave. 
All the imaginary parts of the $P$ waves are consequently set to zero. 
No $D$ waves are included.
If we include the $D$ waves from the Born terms of photoproduction,
this model would correspond to the one used 
in~\cite{Hornidge:2012ca,Fernandez-Ramirez:2014ffa,Fernandez-Ramirez:2013qqa,Schumann:2015ypa} 
to analyze the experimental data. In this way, we keep the $\mathcal{B}$-model as realistic as possible.
The synthetic data are generated around that solution 
at the same energies and scattering angles as the real data and with the same error bars.

{\bf 2.} 
These synthetic data are then analyzed with the full 46-parameter 
model as defined in the previous section, minimizing the penalized 
$\chi^2_T=\chi^2+P$ 
for different $\lambda$ according to Eq.~(\ref{totalchi}).

\begin{figure}
\begin{center}
\includegraphics[width=0.9\linewidth]{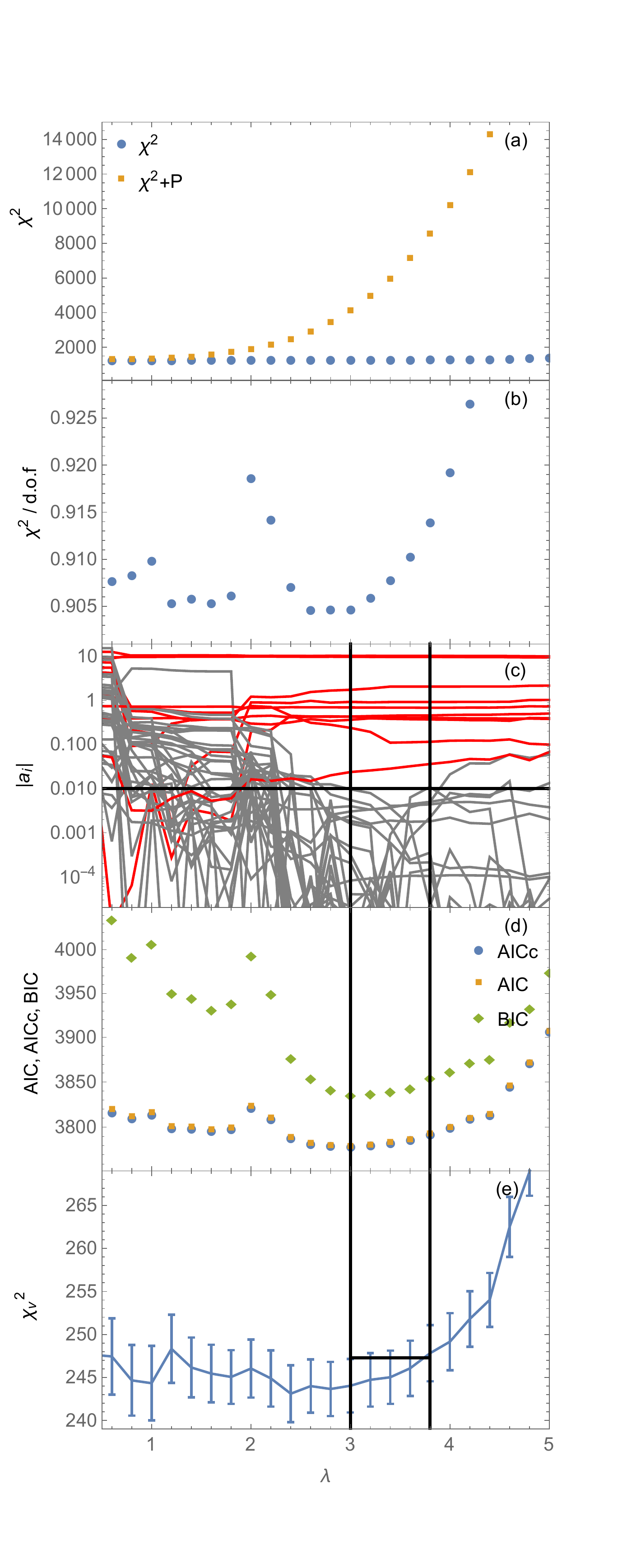}
\end{center}
\caption{
LASSO, information criteria, and cross validation for the $\mathcal{B}$-model. 
(a) Total $\chi_T^2(\lambda)=\chi^2(\lambda)+P(\lambda)$ with penalty $P$ (orange points) 
and $\chi^2(\lambda)$ from data alone (blue points). 
(b) The $\chi^2$/d.o.f. with the number of parameters dynamically updated for each $\lambda$. 
(c) Absolute value of the parameters $a_i$ as function of $\lambda$ in a logarithmic scale. 
The red lines indicate the finally chosen parameters, the gray lines show the unnecessary parameters. 
(d)  AIC, AICc and BIC. 
(e) The cross-validation $\chi^2_V$. See text for further explanations.}
\label{fig:lasso1}
\end{figure}

In Fig.~\ref{fig:lasso1}(a) 
the total $\chi^2_T$ and the contribution from data alone ($\chi^2$) is indicated. 
The difference is given by the penalty $P$. 
Both curves rise as $\lambda$ increases, and therefore, as discussed, 
one needs additional criteria to determine the optimal value, or range of values, for $\lambda$. 
In Fig.~\ref{fig:lasso1}(b) the $\chi^2$ per degree of freedom
$\chi^2$/d.o.f. is shown. 
It exhibits a minimum at around $\lambda=3$ which already provides 
an initial impression about where to look for the simplest model. 
Yet, note that a simple Pearson's $\chi^2$ test at 90\% lower CL would rule out all fits up to $\lambda\approx 4$ as overfits,
such that it is difficult to justify the use of the $\chi^2$/d.o.f. itself as a tool to determine the optimal value for $\lambda$.  
 
The $\lambda$-dependence of the fit parameters $a_i$ is shown in Fig.~\ref{fig:lasso1}(c). 
We have chosen here a logarithmic scale; on a linear scale, 
it becomes obvious that the parameters effectively approach zero once the penalization is large enough. 
Yet, the picture shows that, sometimes, parameters that had died out can in principle reappear at larger values of $\lambda$. 
Figure \ref{fig:lasso1}(c) suggests that for $\lambda\ >3$ many parameters are effectively zero 
and the optimal value for $\lambda$ is expected in that region. 
Anticipating the final result, the figure shows effectively non-zero parameters 
in red while effectively zero parameters are shown in gray. 
The horizontal line indicates the cut-off below which a parameter is counted as zero. 
It is remarkable to observe the parameters drop by three orders of magnitude because it means that LASSO 
is also capable of disentangling the extreme correlations between parameters present at $\lambda=0$, 
as the unnaturally large parameter values at $\lambda=0$ indicate.

As Fig.~\ref{fig:lasso1}(d) shows, the AIC(c) and BIC criteria confirm this picture quantitatively, 
with all three considered criteria exhibiting a minimum at the same $\lambda=3$. 
The BIC shows the cleanest signal as the region to the left of the minimum shows the steepest slope. 
The cross-validation $\chi^2_V$ shown in Fig.~\ref{fig:lasso1}(e), 
obtained through 5-fold validation~\cite{Lasso, ISL}, exhibits a broad, 
shallow minimum from $\lambda\approx 2.5$ through $\lambda\approx 3.5$
which does not very well determine the optimal $\lambda$. 
To have an impression of what the simplest allowed model according to the 
$1$-$\sigma$ rule~\cite{Lasso, ISL} could be, 
we can continue the upper end of the error bar, at the optimal $\lambda=3$ found before, 
horizontally until it intersects with the central value of $\chi^2_V$ at $\lambda=3.8$. 
This indicates the simplest model compatible with cross validation within errors. 
Combining the findings from the AIC(c), BIC and cross validation, 
vertical lines at $\lambda=3$ and $\lambda=3.8$ enclose a region of optimal $\lambda$. 
If we return now to Fig.~\ref{fig:lasso1} (c), it becomes apparent that in that region 
the number of effective non-zero parameters indeed does not change, 
meaning that the precise value of $\lambda$ to choose the simplest model does not matter as long as it is within that range.

{\bf 3.} 
Although trivial, the last step consists in setting all parameters 
that have been ruled out by LASSO
to exactly zero 
and refit the model with the remaining parameters to the synthetic data. 
In summary, the LASSO reduces a model with 46 parameters to a simpler 
one with 10 parameters, remarkably close to the true number of parameters (9). 

\subsubsection{Discussion}
\label{sec:discussion1}

\begin{figure}
\begin{center}
\includegraphics[width=0.99\linewidth]{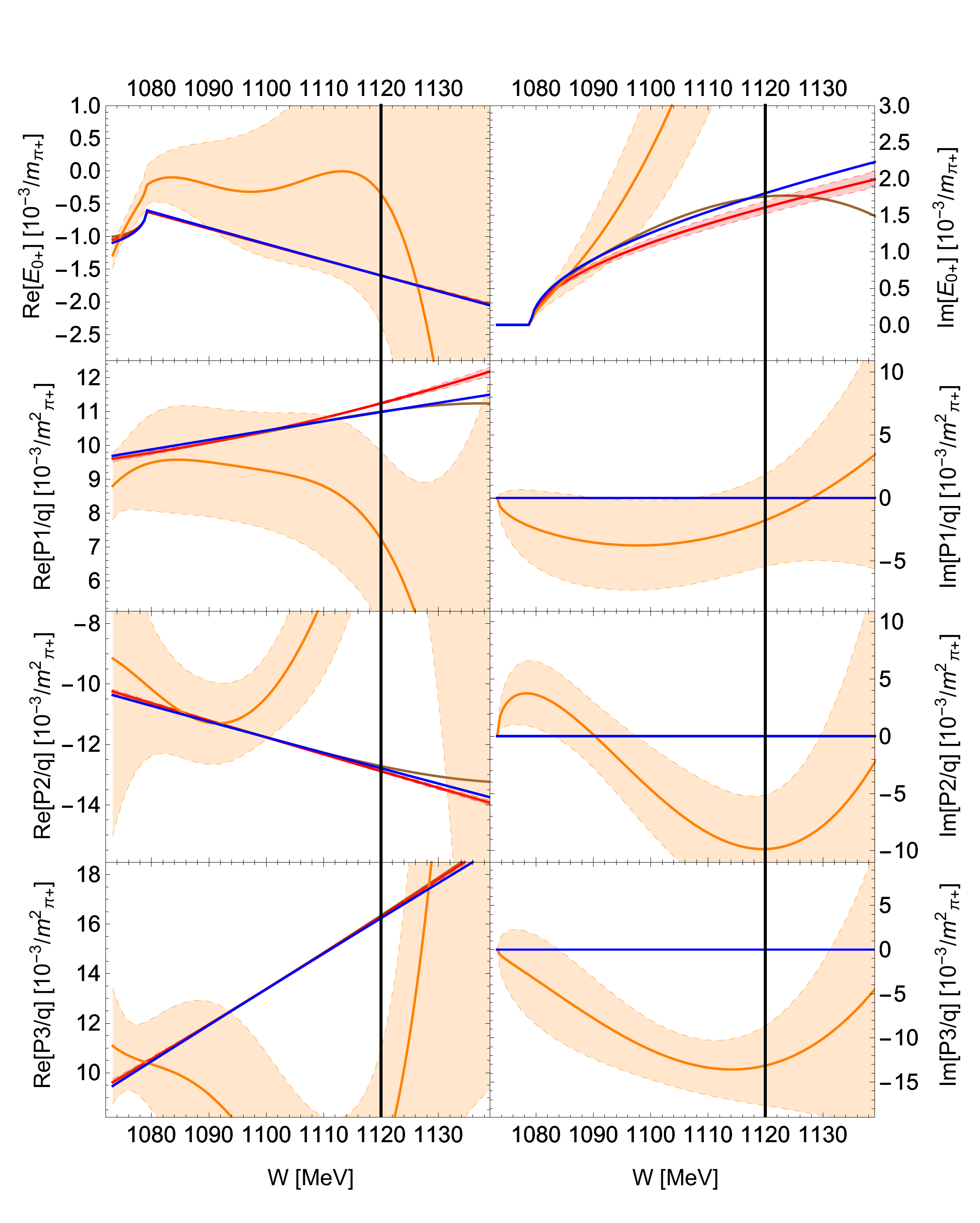}
\end{center}
\caption{
$S$ and $P$-wave partial waves 
from the fits to the $\mathcal{B}$-model 
($D$ waves are not shown because the result obtained was identically zero as expected). 
The blue lines show the $\mathcal{B}$-model used to generate synthetic data (benchmark solution). 
The orange lines and bands show the unconstrained 46-parameter fit with zero penalty $\lambda=0$. 
The brown lines show the fit at optimal $\lambda=3$. 
Removing the parameters that are effectively zero and refitting produces the final result indicated with red lines and uncertainty bands. 
The vertical lines at $W=1120$~MeV indicate the upper limit of fitted data. 
Note that for $\text{Im}\, P_i$, all but the unconstrained (orange) results for $\lambda=0$, are identically zero. 
Also in other cases the curves lie on top 
of each other such that only the blue line is visible.
Uncertainties are computed using  bootstrap.
}
\label{fig:multis1}
\end{figure}

In Fig.~\ref{fig:multis1}, the different steps in the model selection process are illustrated. 
The blue lines show the partial waves used to generate the synthetic data and provide, thus, 
the benchmark solution (${\cal B}$-model) to be reproduced. 
The over-fit at $\lambda=0$ is indicated with the orange lines and bands. All uncertainty bands have been obtained by bootstrap 
as explained in Sec.~\ref{sec:bootstrap}.
Clearly, at the cost of fitting fluctuations, the benchmark solution is missed, even within the large uncertainties. 
Note that this is a 46-parameter fit to altogether 1373 data points, meaning that from these numbers 
it is per se not obvious that we have an overfit. 
However, the $\chi^2$ of this fit ($\chi^2=1232$) is below the lower 95\% confidence limit of Pearson's 
$\chi^2$ test at $\chi^2=1243$, indicating overfitting.

We stress here the well-known defect of the rule-of-thumb, that a $\chi^2$/d.o.f. $\approx 1$ indicates a good fit. 
For a few points, a $\chi^2$/d.o.f. significantly larger than one might be perfectly acceptable, while a $\chi^2$/d.o.f.$=1.1$ 
for a fit to 10,000 data points is very bad. 
Pearson's $\chi^2$ test gives here a better answer with respect to under-fitting.

Finally, the 46-parameter fit has also greatly reduced predictability. 
Here, we have only included data up to a cut-off of $W\le 1120$~MeV as indicated with the vertical lines in the figure. 
As expected, beyond that value, the fit shows large oscillations and an uncontrolled increase of the uncertainties (orange bands).

The fit at $\lambda=3$ is shown with the brown lines in Fig.~\ref{fig:multis1} 
(as discussed before, we could have taken any value up to $\lambda=3.8$). 
Removing all parameters that are effectively zero and refitting the remaining 10 parameters, 
the final solution is indicated with the red lines and uncertainty bands. 
It remarkably well reproduces the benchmark solution (blue lines) 
and even produces reasonable and well-constrained partial waves 
beyond the range of fitted data (indicated with vertical lines). 

The one additional parameter of the 10-parameter fit, compared to
the benchmark solution (9 parameters), is given by $a_1$ in Eq.~(\ref{eq:ime0p}). Yet, 
as Fig.~\ref{fig:multis1} shows, the benchmark solution for Im~$E_{0+}$ is 
rather well reproduced. At higher energies, the benchmark solution is 
just slightly outside the red narrow uncertainty band indicating the 68\% confidence region.

It is also remarkable that in the simplest model LASSO is capable of setting all $D$ waves to zero, 
which are finite in the over-parametrized model ($\lambda=0$) but absent in the benchmark solution. 
Furthermore, the imaginary parts of all $P$ waves are found to vanish (cf. Fig.~\ref{fig:multis1}), 
in agreement with the benchmark solution. 

One should note that in the fit to photoproduction data there
is an overall-phase problem that is usually solved by fixing the phase of one multipole. 
Here, the full 46-parameter model indeed contains only real $D$ waves while the phase of all other multipoles are undetermined. 
Yet, those $D$ waves are very small (at $\lambda=0$) such that the overall phase problem is in fact not very well fixed. 
Part of the wide error bands for the $\lambda=0$ case can, 
thus, be attributed to a rather uncontrolled variation of the overall phase, 
leading to a large reshuffling between real and imaginary parts and causing the very poorly determined multipoles.

Finally, one can discuss other goodness-of-fit criteria beyond Pearson's $\chi^2$ test. 
For example, it is common to find a best fit by combining the $\chi^2$ test --that is rather restrictive to under-fitting 
but more tolerant to over-fitting-- with the $F$ test that determines, for nested models, 
if a more complex model leads to a significant improvement. 
For that, the $\chi^2$s of two models with $k$ (model 1) and $m+k$ (model 2) parameters fitted 
to $n$ data points are compared. 
If the true values of the $m$ extra parameters of the more complex model vanish, it can be shown that 
\be
y=\frac{(\chi^2_1-\chi^2_2)/k}{\chi^2_2/(n-m-k)}
\ee
is $F(k,n-m-k)$ distributed~\cite{FTest}. 
For the explicit form of the $F$-distribution, see, e.g., Ref.~\cite{freund}. 
A value of $y$ beyond a chosen CL limit thus indicates 
that the more complex model 2 is significantly better than the simpler model 1 
(which cannot be judged from the $\chi^2$ values alone). 
Here, we find that $y$ obtained from the 46-parameter fit and the optimal (simplest) 10-parameter fit is $y=1.64$ 
which is below the 90\% CL interval ending at $y=2.63$, 
indicating that the overfit is indeed not significantly better than the simplest fit.

\begin{figure}
\begin{center}
\includegraphics[width=0.9\linewidth]{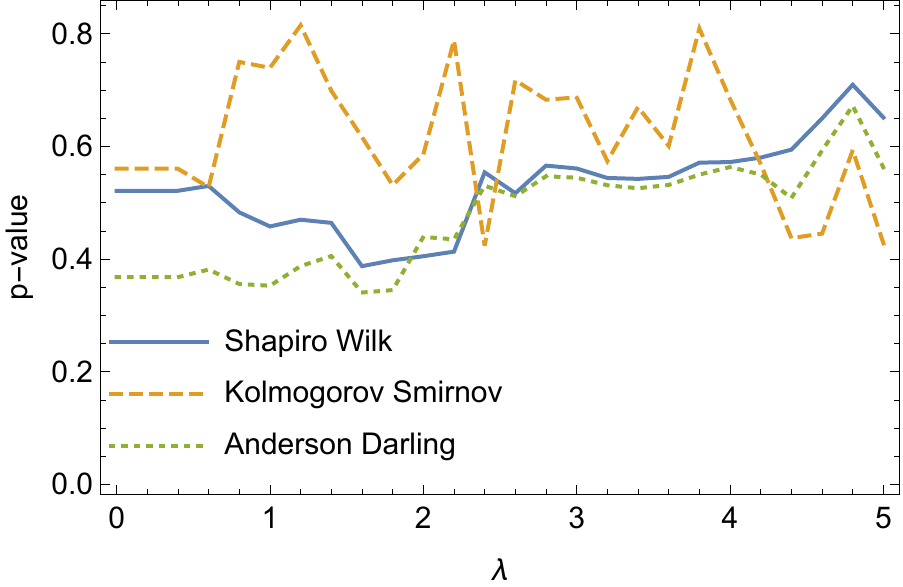}
\end{center}
\caption{Different goodness-of-fit criteria as a function of the penalty parameter $\lambda$.}
\label{fig:criteria}
\end{figure}

Other common goodness-of-fit-criteria are the 
Shapiro-Wilk~\cite{Shapiro}, 
Anderson-Darling~\cite{AD1952, Stephens1974}, 
and the Kolmogorov-Smirnov~\cite{Kolmogorov,Smirnov,Stephens1974} tests. 
They provide means to compare a sampled probability distribution against a theoretically expected one. 
In particular, they can be used to test fit residuals against their expected Normal $N[0,1]$ distribution. 
This applies also for weighted data as considered here as long as the fit residuals 
are divided by the respective experimental (statistical) uncertainty. 
In particular, these tests are by construction more sensitive than Pearson's $\chi^2$ test because, on one hand, 
they are sensitive to the sign of the fit residual and not only its square, and, more importantly, 
they test the entire distribution instead of only the bulk property given by the sum of individual $\chi^2$.
The $p$-values of the three tests are shown in Fig.~\ref{fig:criteria} as a function of $\lambda$.

For all considered values of $\lambda$ the $p$-values are high; usually, fits with $p>0.05$ are considered acceptable. 
For small values of $\lambda$ the tests score high although one is in the region of over-fitting. 
In the region of optimal $\lambda$ between $\lambda\approx 3$ and $\lambda\approx 4$, 
the $p$ values are also consistently $p>0.4$; however, there is no clear trend that would allow one to use these criteria
themselves to find an optimal value for $\lambda$.

\subsection{LASSO for real data} \label{sec:realdata}

\begin{figure}
\begin{center}
\includegraphics[width=0.95\linewidth]{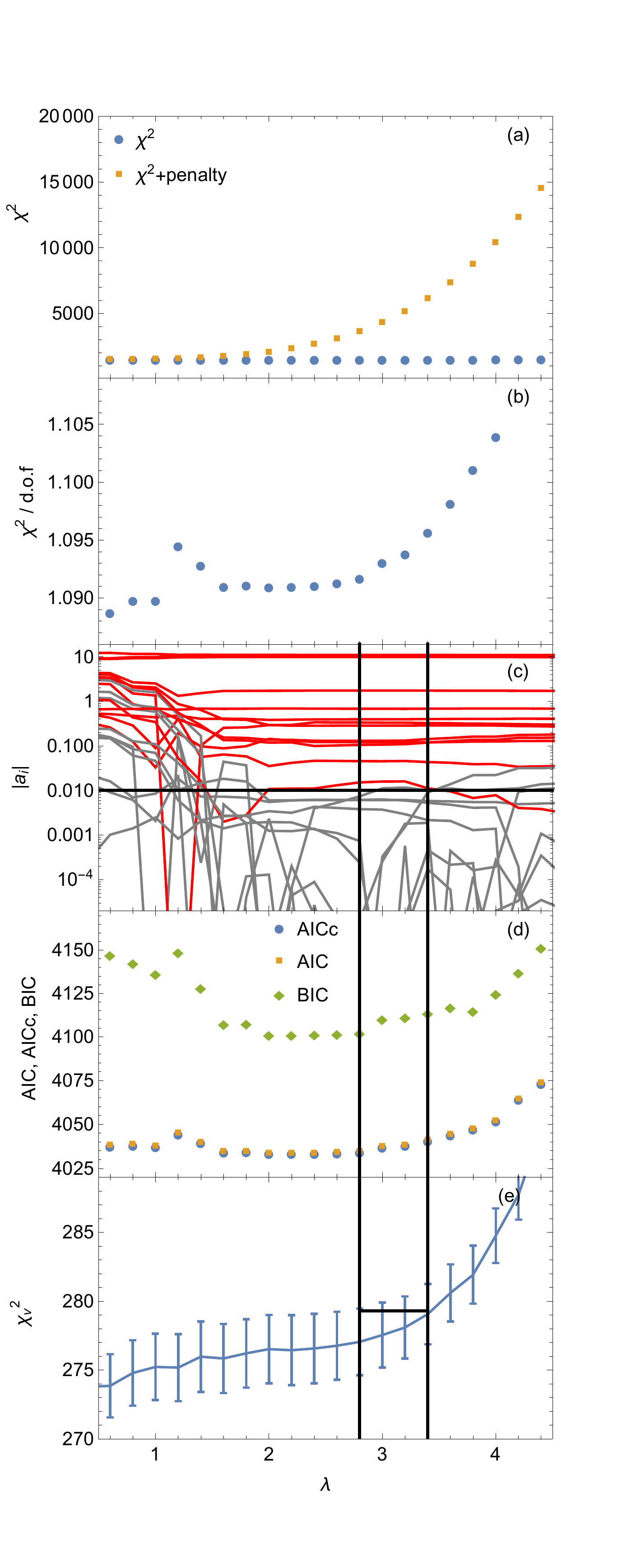}
\end{center}
\caption{The $\lambda$ dependence of the same quantities as discussed in Fig.~\ref{fig:lasso1} 
but for the situation with actual data from experiment.
}
\label{fig:lasso2}
\end{figure}

For the reaction $\gamma p\to\pi^0p$,
precise low-energy data for the differential cross section $d\sigma/d\Omega$ 
and photon beam asymmetry $\Sigma$ are available from MAMI~\cite{Hornidge:2012ca}; 
for earlier measurements see Refs.~\cite{Schmidt:2001vg, Bergstrom:1997jc, Fuchs:1996ja, Beck:1990da}. 
The target polarization differential cross section $d\sigma_T/d\Omega$ 
has been measured recently at MAMI as well~\cite{Schumann:2015ypa}.
At energies beyond those considered here, 
these observables have also been remeasured with unprecedented precision~\cite{Adlarson:2015byy, Annand:2016ppc}.
Electroproduction of $\pi^0$ mesons close to threshold has been recently measured by CLAS~\cite{Chirapatpimol:2015ftl}.
We analyze here the data from \cite{Hornidge:2012ca, Schumann:2015ypa} for $d\sigma/d\Omega$, $\Sigma$, 
and $d\sigma_T/d\Omega$ from the $\pi^0p$ threshold up to $W=1120$~MeV. 

For the analysis of real data, we have slightly reduced the model space used for the $\mathcal{B}$-model.
Before, the $D$ waves were kept real to fix the overall-phase ambiguity. 
However, this is not effective at low energies where $D$ waves are extremely small as discussed. 
Then, for complex $S$ and $P$ waves, the ambiguity reappears at low energies. 
We have therefore set all imaginary parts of the $P$ waves to zero. 
For the $D$ waves themselves, we have kept them fixed at the real values given by the Born terms of photoproduction. 
This is the same procedure as in \cite{Hornidge:2012ca}. 
In total, the number of available parameters is 23. 
Note that none of these changes have to do with the model selection methods discussed here. 
The former are additional constraints imposed from other considerations, 
because we aim at a result that is comparable with the analysis of \cite{Hornidge:2012ca}.

Performing the LASSO scan as in the case of the $\mathcal{B}$-model, 
in combination with the information theory criteria and cross validation, we obtain the results shown in Fig.~\ref{fig:lasso2}. 
The fit is compatible with Pearson's $\chi^2$ test for any $\lambda$ shown. 
The comparison of Fig.~\ref{fig:lasso2} with Fig.~\ref{fig:lasso1} shows quite similar features.
Again, the BIC delivers the most pronounced minimum, or minimal plateau. 
We choose an optimal $\lambda=2.8$ which coincides with the minimum of the BIC which is also the last point of the plateau. 
For the cross validation $\chi^2_V$ there is no minimum at all in this case. 
Only an upper bound for $\lambda$ can be determined by proceeding as before for the $\mathcal{B}$-model, 
i.e., continuing the upper end of the error bar at $\lambda=2.8$ to the right as indicated (in analogy to the 1-$\sigma$ rule). 
This leads to a maximal value of $\lambda=3.4$ that is compatible within errors. 
As Fig.~\ref{fig:lasso2}(c) shows, a parameter rises briefly above the cut-off threshold for values of $2.8<\lambda<3.4$. 
We do not see this as a problem but rather as a feature of the LASSO method demonstrating that the method indeed 
scans a large classes of models, and even rehabilitates parameters that were found to be zero for smaller $\lambda$. 
From the discussion it becomes clear that in this case it is not possible to fix the precise number of parameters without doubts; 
yet, we have seen before for the $\mathcal{B}$-model that it is possible to determine that number approximately. 
Overall, the number of model parameters is reduced from 23 to 13  (in the $\mathcal{B}$-model: from 46 to 10). 

\subsubsection{Discussion}\label{sec:realdatadiscussion}
\begin{figure}
\begin{center}
\includegraphics[width=0.99\linewidth]{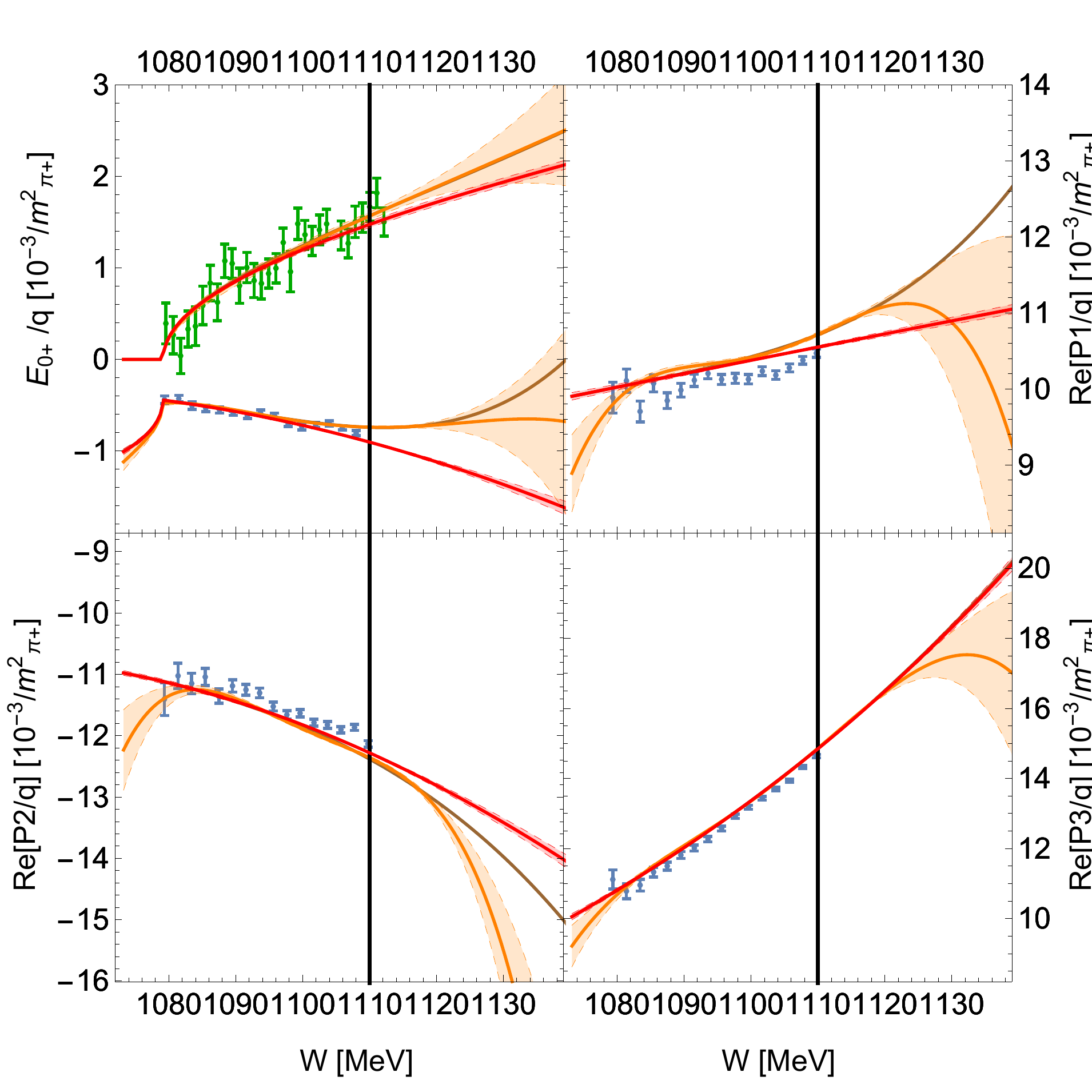}
\end{center}
\caption{$S$- and $P$-wave partial waves
from the analysis of real data. 
Labeling of the curves and uncertainty bands as in Fig.~\ref{fig:multis1}. 
The simplest model is indicated with the red lines and bands. 
The blue circles show the single-energy solution of \cite{Hornidge:2012ca} 
for the real parts of $S$- and $P$-wave partial waves. 
The green triangles indicate the single-energy extraction of $\text{Im}\,E_{0+}$
from \cite{Schumann:2015ypa}.
Uncertainties are computed using  bootstrap.
}
\label{fig:multis2}
\end{figure}

The partial waves are shown in Fig.~\ref{fig:multis2}. 
They share similar properties as observed for the $\mathcal{B}$-model in Fig.~\ref{fig:multis1}.

For the unconstrained, $\lambda=0$ case with 23 parameters, 
the uncertainty band in the region of fitted data is much narrower than in the $\lambda=0$ case for the $\mathcal{B}$-model. 
This may be partly explained by the smaller maximal number of parameters available; 
another reason is the discussed overall-phase ambiguity which is avoided here by allowing only for real $P$-wave multipoles. 

For the intermediate range of energies, the solutions and uncertainty bands are sometimes difficult to distinguish in Fig.~\ref{fig:multis2}. 
A closer inspection reveals that
the uncertainty bands of the 23-parameter fit are still wider than the ones of the 13-parameter simplest model.
We also observe largely widened error bands for the unconstrained fit at very low energies and beyond 
the fitted region (vertical bands) indicating the reduced predictability of the $\lambda=0$ fit. 
The simplest model (red lines and bands) shows good qualitative agreement with the SE 
solution from \cite{Hornidge:2012ca} for the real parts of the partial waves, 
although they do not coincide perfectly. 
There are differences such that here we allow for a variation of the cusp parameter
at threshold, $\beta_0=a_0 \: m^{-1}_{\pi^+}$ in Eq.~(\ref{eq:ime0p}),
which was held fixed in \cite{Hornidge:2012ca} and, second, we include here the new 
$d\sigma_T/d\Omega$ data from \cite{Schumann:2015ypa}. 
The optimal (simplest) solution is also very close to the result of \cite{Ronchen:2014cna} (not shown here) 
that includes isospin breaking in a similar way as implemented here, 
fulfills Fermi-Watson's theorem and extends up to the resonance region. 

The imaginary part of $E_{0+}$ of the simplest model agrees well with the SE extraction performed 
in \cite{Schumann:2015ypa} (green data points) as Fig.~\ref{fig:multis2} shows.
For the cusp parameter at threshold we obtain
$\beta_0=\left( 2.41\pm 0.05 \right) \cdot 10^{-3}\,m_{\pi^+}^{-1}$.
This agrees with the value from \cite{Schumann:2015ypa}: 
$\beta_0= \left( 2.2\pm 0.2\text{[stat.]}\pm 0.6\text{[syst.]} \right)\cdot 10^{-3}\,m_{\pi^+}^{-1}$. 
Note, however, the very different size of the statistical uncertainty which comes here from a global fit to all data while 
in \cite{Schumann:2015ypa} it was obtained from a two-parameter fit to the SE solutions shown in  Fig.~\ref{fig:multis2} with the green triangles. 
Note that the systematic uncertainties of $\beta_0$ are estimated to be $0.6$ in the above units~\cite{Schumann:2015ypa} 
which is larger than the statistical ones. 
This shows that systematic effects are not negligible, 
but we have refrained from analyzing these effects here because the focus of this paper is different. 
It should be mentioned that the values for $\beta_0$ found here and in \cite{Schumann:2015ypa} are smaller than the value 
of $\beta_0=3.35\cdot 10^{-3}\,m_{\pi^+}^{-1}$ obtained in ChPT calculations~\cite{Hoferichter:2009ez, Baru:2011bw}
using pionic atoms data. 

\begin{figure*}
\begin{center}
\includegraphics[width=0.9\linewidth]{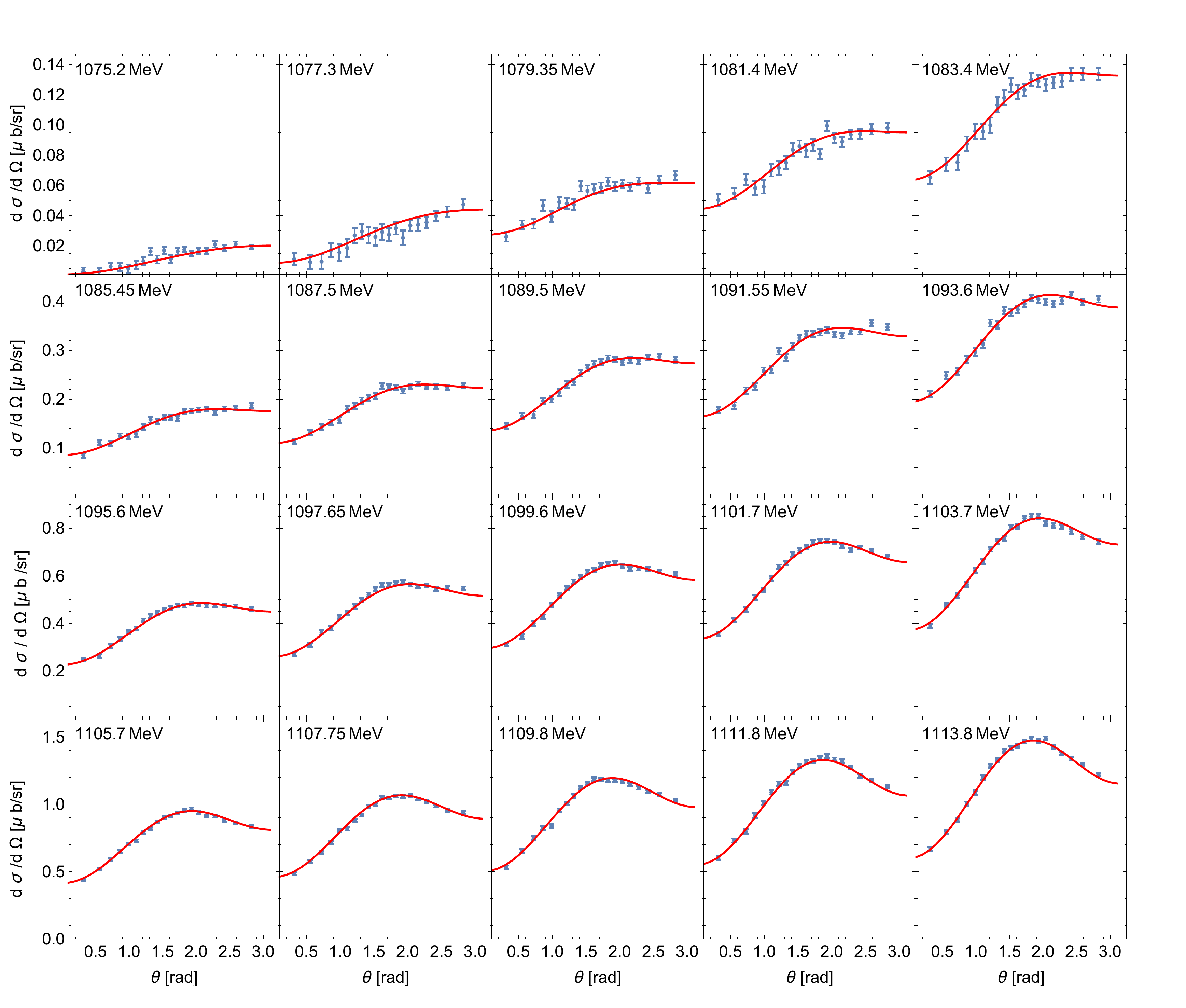}
\end{center}
\caption{Differential cross section. Data from \cite{Hornidge:2012ca}. 
The labels in the plot indicate the scattering energy $W$ in MeV. 
The red lines show the central value of the simplest model as determined in this study 
(indicated with the red lines in Fig.~\ref{fig:multis2}). 
}
\label{fig:dsdo}
\end{figure*}

\begin{figure*}
\begin{center}
\includegraphics[width=0.99\linewidth]{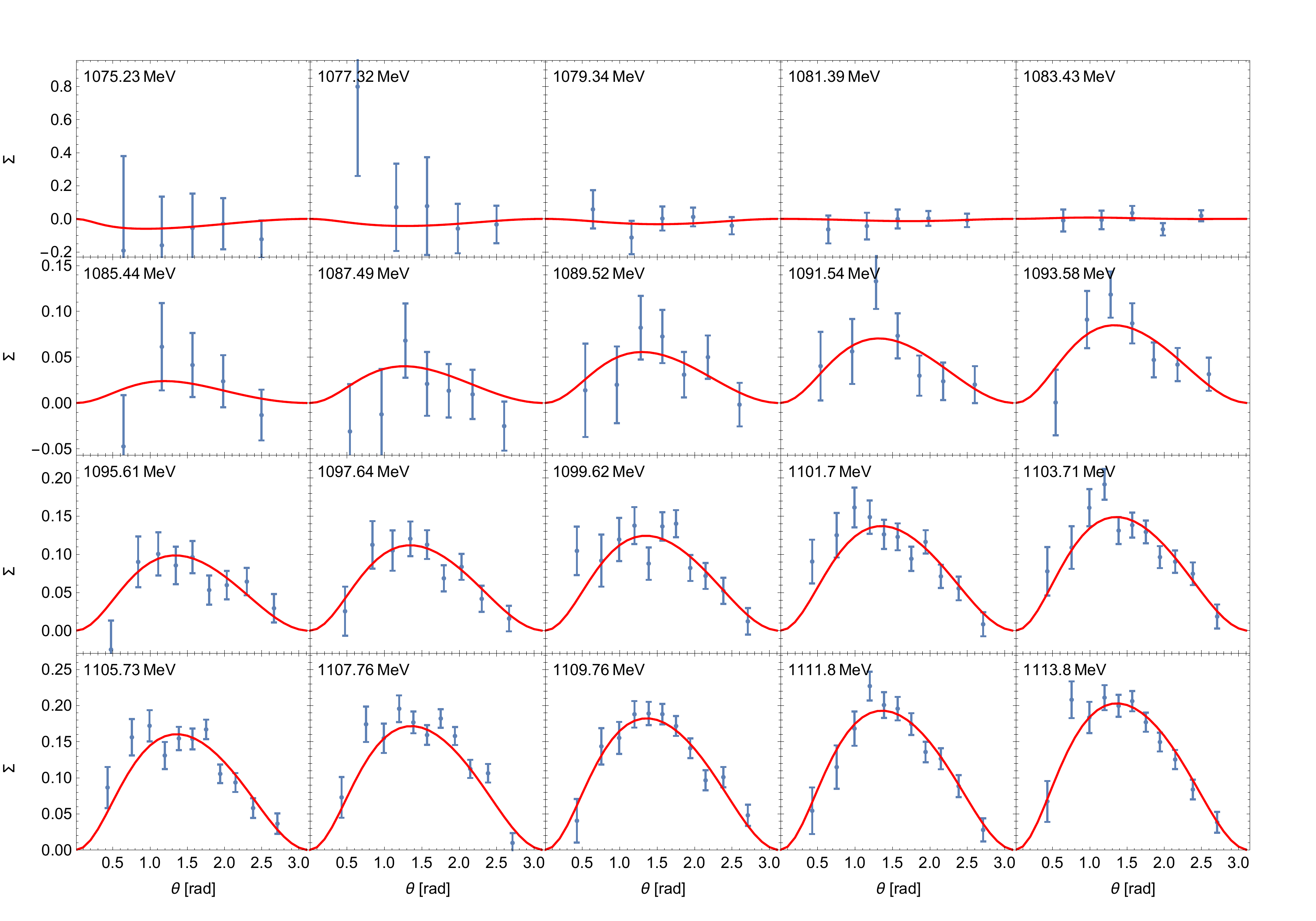}
\end{center}
\caption{Beam asymmetry. Data from \cite{Hornidge:2012ca}. See Fig.~\ref{fig:dsdo} for further description. 
}
\label{fig:Sigma}

\end{figure*}

\begin{figure*}
\begin{center}
\includegraphics[width=0.99\linewidth]{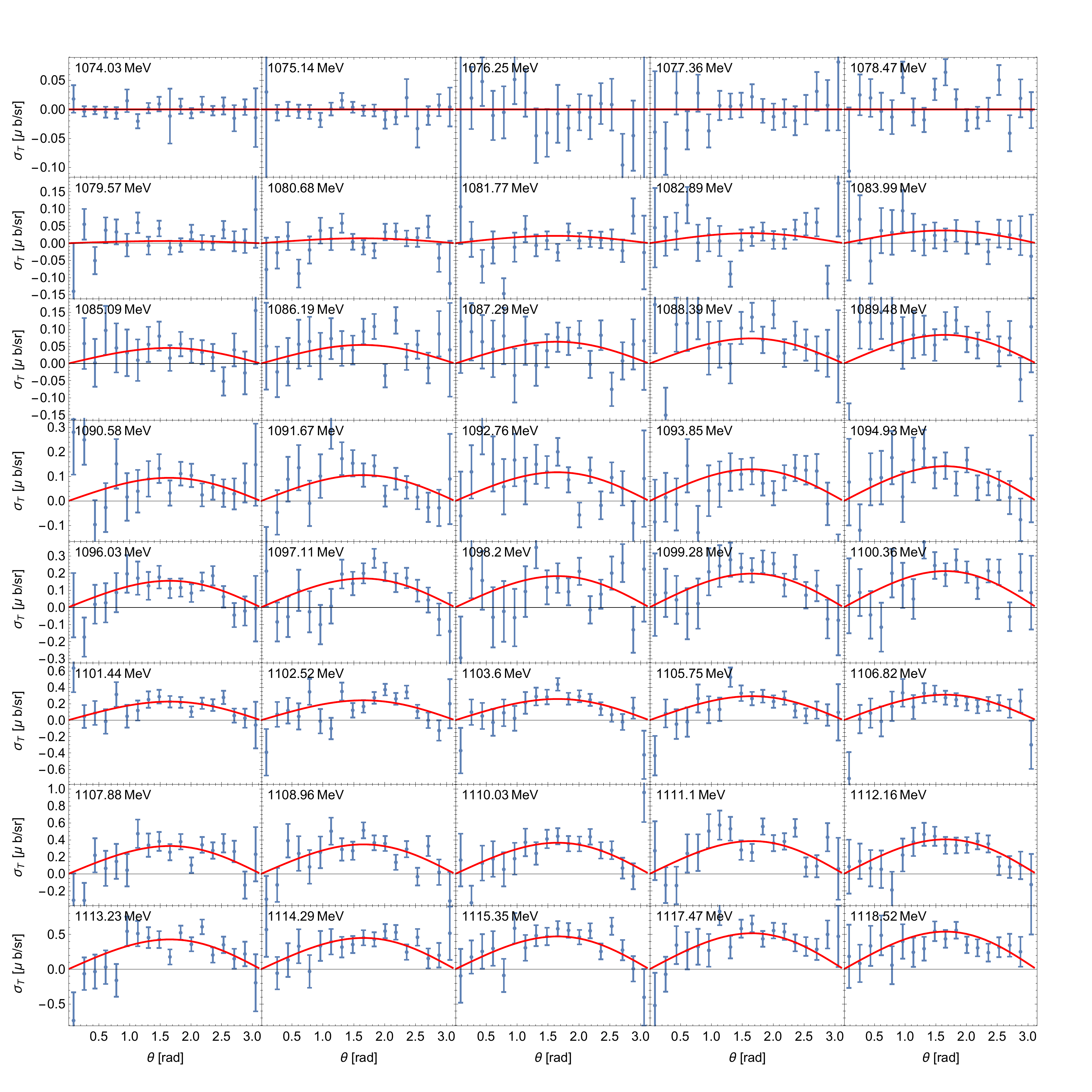}
\end{center}
\caption{Polarized differential cross section $\sigma_T=d\sigma/d\Omega\,T$. 
Data from \cite{Schumann:2015ypa}. See Fig.~\ref{fig:dsdo} for further description. 
}
\label{fig:T}
\end{figure*}

In Figs.~\ref{fig:dsdo}, \ref{fig:Sigma}, and \ref{fig:T}, the experimental data and the best fit result are shown, 
corresponding to the 13 parameter fit indicated with the red lines in Fig.~\ref{fig:multis2}. 
The description of the data is consistently good. 
In particular, the inclusion of the new $d\sigma_T/d\Omega$ data 
does not have much influence in the multipoles other than reducing their uncertainties.


As a concluding remark it should be mentioned that the AIC and BIC compare the relative quality of models and are, thus, 
also applicable in situations in which no good fit in absolute terms can be achieved 
(e.g., when the achievable $\chi^2$ is too large for all models). 
This robustness is a relevant aspect in the analysis of photoproduction data, 
especially for excited baryons at higher energies. 
There, data from different experiments are sometimes plagued by underestimated systematic uncertainties or even contradictory
(for the selection of consistent data sets see, e.g., Refs.~\cite{Perez:2013jpa, Perez:2014yla}). 
Then, no good $\chi^2$ might be achieved and one has to proceed using relative comparisons of models as proposed here.

\section{Conclusions}
The LASSO provides a tool to scan large classes of models through the ability to set fit parameters to zero. 
Together with criteria from information theory and cross validation, it is possible to select a {\it simplest} 
model with a minimum of fit parameters. 
Here, we concentrate on single-meson photoproduction. 
The methods are, however, not restricted to this case and also applicable, 
e.g., in the context of excited meson production experiments as recently 
discussed by Guegan, Hardin, Stevens, and Williams~\cite{Guegan:2015mea}. 

In the example of pion photoproduction at low energies, 
we show that the LASSO in combination with additional criteria decreases 
the uncertainties of extracted multipoles and increases predictability, as expected. 
First, a benchmark model  ($\mathcal{B}$-model) was considered to show these properties. 
The overall-phase problem was not entirely fixed in that model. 
Still, LASSO showed capable of simplifying this partially ill-defined problem, 
reducing an initial number of 46 fit parameters to 10
and recovering the known benchmark solution with remarkable accuracy and precision. 
In particular, properties of the benchmark solution such as vanishing imaginary parts of the $P$ waves 
and absence of $D$ waves could be detected.  
The considered criteria for the determination of the optimal penalty $\lambda$ --AIC, AICc, and BIC-- 
consistently exhibit minima at the same value of $\lambda$. 
The 1-$\sigma$ rule in cross validation also defines a whole range of $\lambda$ 
to be considered as optimal, and indeed in that range the number of relevant parameters stays constant.

When moving on to the analysis of real data, similar observations could be made. 
The simplest model shows remarkable agreement with single-energy extractions of multipoles from other studies. 
Additionally, these techniques provide more stable and reliable extrapolations of the models for energies above
and below the fitted region.
This feature can be exploited to determine physical observables at threshold, 
e.g. the discussed cusp parameter $\beta_0$.
Besides their usefulness in the analysis of experimental data, the
proposed methods could also be applied in the analysis of lattice QCD
eigenvalues because the infinite-volume extrapolation of coupled-channel systems 
on the lattice necessarily requires an expansion of the
amplitude in energy \cite{Doring:2011nd}, similar to what has been discussed here.
The discussed techniques also  promise for a systematic 
and automatic work flow in the analysis of the excited baryon 
and meson spectra, 
providing a perspective towards finding more conclusive answers in hadron spectroscopy.

\begin{acknowledgments} 
The authors thank Maxim Mai for discussions.
This work is supported by the National Science Foundation 
(PIF grant No. 1415459) and by The George Washington University 
through the Columbian College Facilitating Funds (CCFF).
M.D. is also supported by the U.S. Department of Energy, Office of Science, 
Office of Nuclear Physics under contract DE-AC05-06OR23177 and grant No. DE-SC0016582. 
C.F.-R. is supported in part by CONACYT (Mexico) under grant No. 
251817, by research grant IA101717 from PAPIIT-DGAPA (UNAM) and by Red Tem\'atica 
CONACYT de F\'{\i}sica de Altas Energ\'{\i}as (Red FAE, Mexico).
\end{acknowledgments}


\end{document}